\documentclass[]{spie}  

\usepackage{amsmath,amsfonts,amssymb}
\usepackage{graphicx}
\usepackage[colorlinks=true, allcolors=blue]{hyperref}

\title{A 50\,mK test bench for demonstration of the readout chain of Athena/X-IFU}

\author[a]{Florent Castellani}
\author[a,c,e]{Sophie Beaumont}
\author[a]{François Pajot}
\author[a]{Gilles Roudil}
\author[c,e]{Joseph Adams}
\author[c]{Simon Bandler}
\author[c]{James Chervenak}
\author[b]{Christophe Daniel}
\author[f]{Edward V. Denison}
\author[f]{W. Bertrand Doriese}
\author[a]{Michel Dupieux}
\author[f]{Malcolm Durkin}
\author[b]{Hervé Geoffray}
\author[f]{Gene C. Hilton}
\author[a]{David Murat}
\author[a]{Yann Parot}
\author[b]{Philippe Peille}
\author[d]{Damien Prêle}
\author[a]{Laurent Ravera}
\author[f]{Carl D. Reintsema}
\author[c,e]{Kazuhiro Sakai}
\author[f]{Robert W. Stevens}
\author[f]{Joel N. Ullom}
\author[f]{Leila R. Vale}
\author[c,e]{Nicholas Wakeham}

\affil[a]{IRAP, Université de Toulouse, CNRS, CNES, UPS, 9 Av. du Colonel Roche, 31400 Toulouse, France}
\affil[b]{CNES, 18 Av. Edouard Belin, 31400 Toulouse, France}
\affil[c]{NASA/GSFC, 8800 Greenbelt Rd., Greenbelt, MD 20771, United States}
\affil[d]{APC, 10 Rue Alice Domon et Léonie Duquet, 75013 Paris, France}
\affil[e]{CRESST II/UMBC, Baltimore, MD 21250, United States}
\affil[f]{NIST, Boulder, CO 80305, United States}

\authorinfo{Further author information: (Send correspondence to F. Castellani: florent.castellani@irap.omp.eu)}

\begin{document} 
\maketitle

\begin{abstract}
The X-IFU (X-ray Integral Field Unit) onboard the large ESA mission Athena (Advanced Telescope for High ENergy Astrophysics), planned to be launched in the mid 2030s, will be a cryogenic X-ray imaging spectrometer operating at 55\,mK. It will provide unprecedented spatially resolved high-resolution spectroscopy (2.5 eV FWHM up to 7 keV) in the 0.2-12 keV energy range thanks to its array of TES (Transition Edge Sensors) microcalorimeters of more than 2k pixel. The detection chain of the instrument is developed by an international collaboration: the detector array by NASA/GSFC, the cold electronics by NIST, the cold amplifier by VTT, the WFEE (Warm Front-End Electronics) by APC, the DRE (Digital Readout Electronics) by IRAP and a focal plane assembly by SRON. To assess the operation of the complete readout chain of the X-IFU, a 50\,mK test bench based on a kilo-pixel array of microcalorimeters from NASA/GSFC has been developed at IRAP in collaboration with CNES. Validation of the test bench has been performed with an intermediate detection chain entirely from NIST and Goddard. Next planned activities include the integration of DRE and WFEE prototypes in order to perform an end-to-end demonstration of a complete X-IFU detection chain. 
\end{abstract}


\section{INTRODUCTION}
\label{sec:intro}  

Athena (Advanced Telescope for High-Energy Astrophysics)\cite{Nandra2013} is the ESA second large mission of the Cosmic Vision science program, dedicated to the study of the Hot and Energetic Universe. This mission will address two fundamental questions: how baryonic matter assembles into the large-scale structures we see today and how supermassive black holes grow and shape the observable Universe.

The 12 meter focal length telescope is scheduled for launch in the mid 2030's, by the future Ariane 6 European rocket, into a Lagrange point L1 orbit between the Sun and the Earth. The movable silicon pore optics (SPO)\cite{Barriere2022} X-ray mirror will be able to focus photons on two different instruments situated at the focal plane:
\begin{itemize}
    \item the \textbf{Wide Field Imager (WFI)}\cite{Meidinger2017} optimised for surveys;
    \item the \textbf{X-ray Integral Field Unit (X-IFU)}\cite{Barret2018} optimised for spatially resolved high resolution spectroscopy.
\end{itemize}

The X-IFU is an imaging high-resolution spectrometer based on an array of 2376 transition-edge-sensor (TES) microcalorimeters that are cooled to a bath temperature of 55 mK.  The X-IFU TESs achieve energy resolution better than 2.5\,eV for X-ray energies up to 7\,keV, which is more than a factor of 50 better than the energy resolution of traditional solid-state X-ray imagers. It operates in the soft X-ray energy band, between 0.2 and 12\,keV, and the pixel size is 5" in a hexagonal field-of-view of 5' equivalent diameter\cite{Pajot2018}.

The X-IFU is built under the responsibility of IRAP and CNES by a consortium of 11 European countries plus USA and Japan. The instrument is currently at the end of its preliminary definition Phase (Phase B). In this context, we need to assess the performance of the instrument, and in particular to validate the operation of a complete prototype readout chain.

\section{THE CNES/IRAP CRYOGENIC TEST BENCH: ELSA}
\label{sec:elsa}

\begin{figure} [ht]
   \begin{center}
   \begin{tabular}{c} 
   \includegraphics[height=9cm]{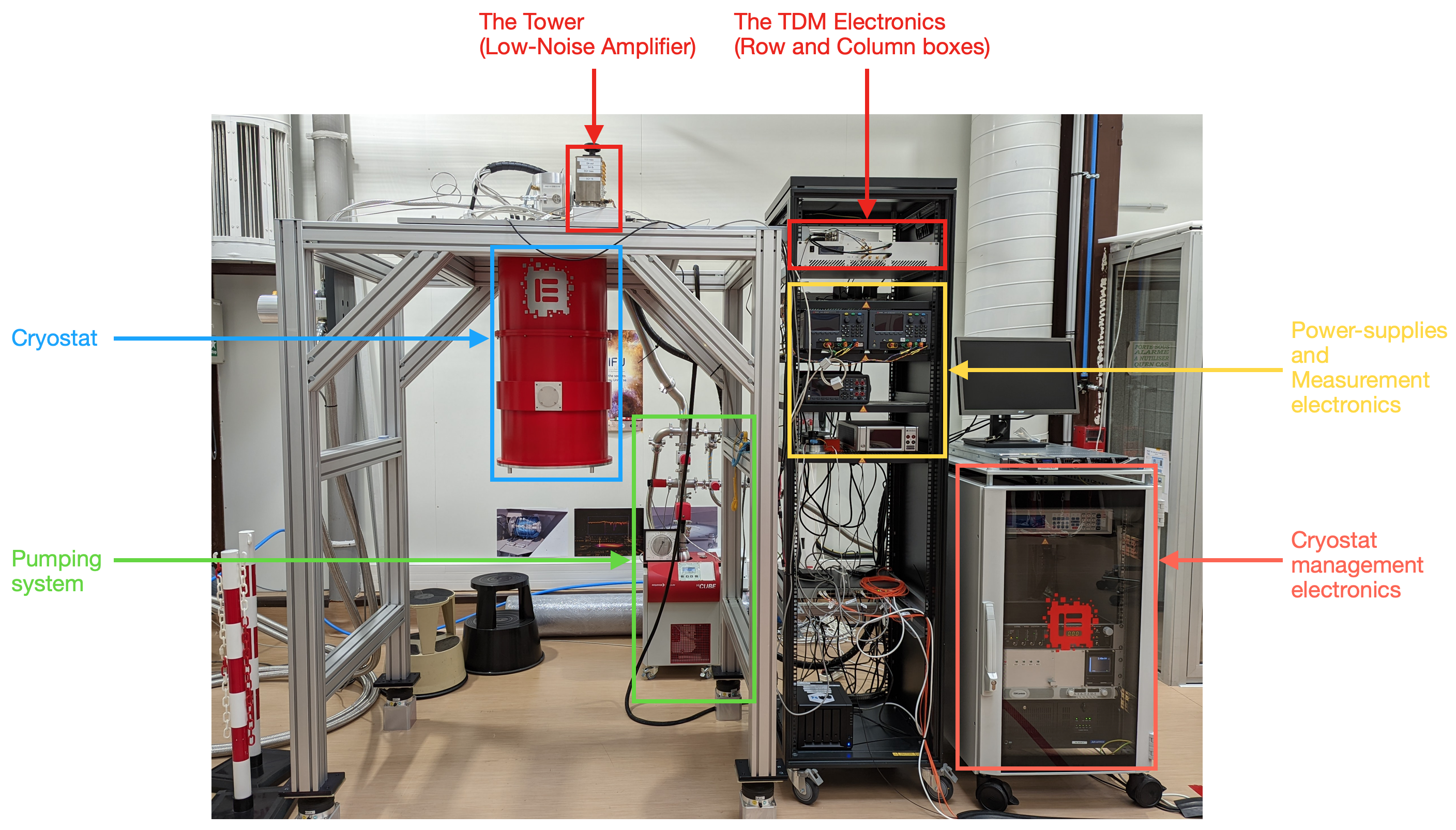}
   \end{tabular}
   \end{center}
   \caption[bench] 
   { \label{fig:bench} 
The CNES-IRAP 50\,mK cryogenic test bench: Elsa}
   \end{figure}

To demonstrate the operation of the warm electronics blocs of the X-IFU readout chain with representative cold electronics and microcalorimeters, a cryogenic test bench has been developed at IRAP in collaboration with CNES, ``Elsa" (Fig.~\ref{fig:bench}).

\subsection{Commercial cryostat from Entropy GmbH}
\label{sec:cryo}

The Entropy GmbH (L-series) cryostat\footnote{\url{https://www.entropy-cryogenics.com/}} is based on a double stage pulse tube refrigerator providing cooling power at 50\,K  and 3\,K, where a double stage Adiabatic Demagnetisation Refrigerator (ADR) is located, with a gadolinium gallium garnet (GGG) stage at 500\,mK and an ferric aluminum alum (FAA) stage at 50\,mK. The thermometry of the cryostat is operated by an AVS 47-B resistance bridge, to control the different stages temperature. An additional LakeShore 372 resistance bridge has been installed, dedicated to thermal regulation of the FAA stage during experiments.

\subsection{Cold part of detection chain}
\label{sec:snout}

\begin{figure} [ht]
   \begin{center}
   \begin{tabular}{c} 
   \includegraphics[height=8cm]{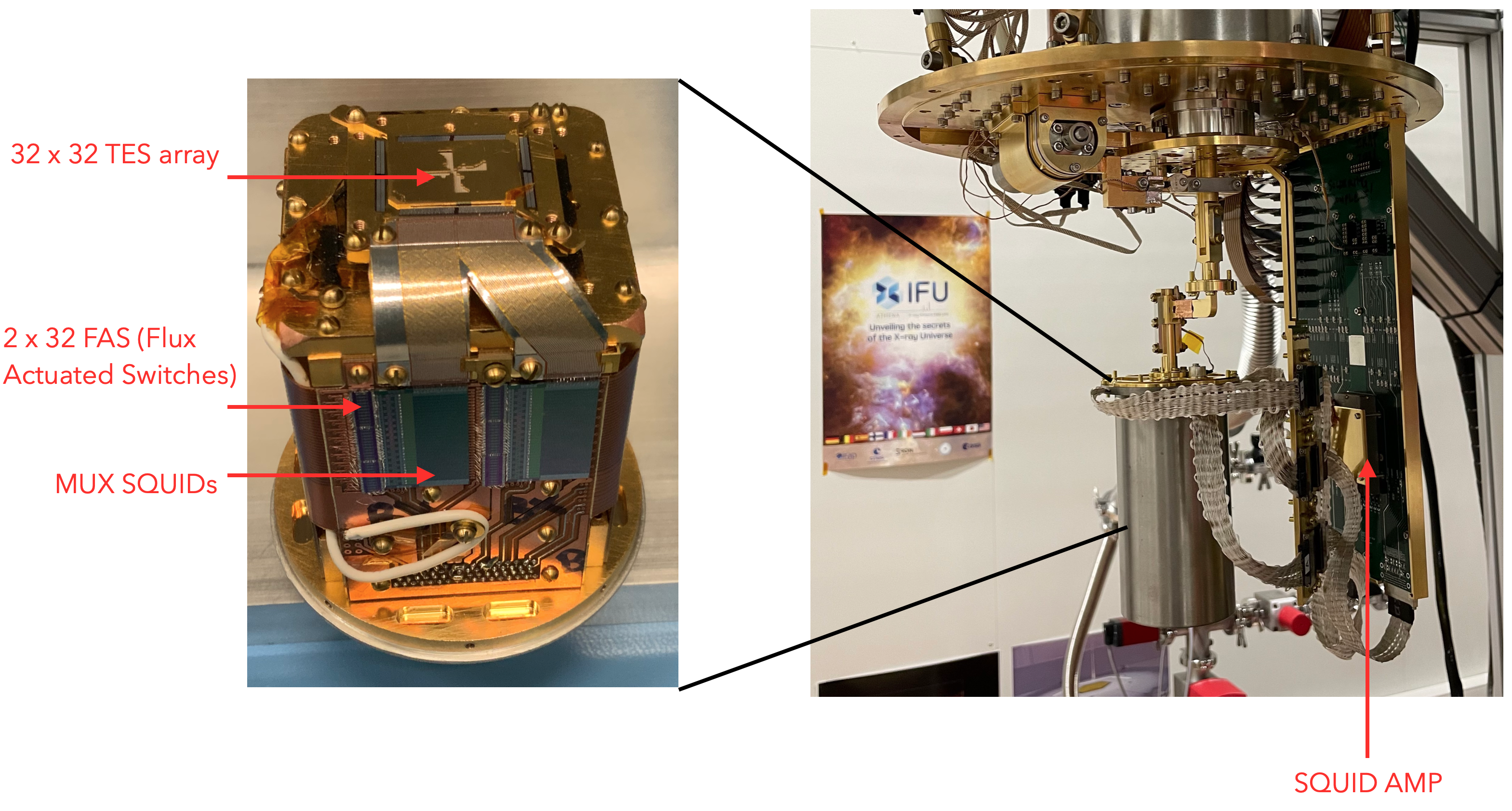}
   \end{tabular}
   \end{center}
   \caption[snout] 
   { \label{fig:snout} 
(Left) The snout is composed of a 1024 TES array with its associated cold readout electronics. (Right) The snout is placed in the niobium shield and connected to the 50\,mK stage. Superconducting looms insure the signals connection to the 500\,mK terminator card and then to the 3\,K cold electronics.}
   \end{figure}

A focal plane assembly from NIST and GSFC, called ``snout"\cite{Doriese2017}, is attached to the 50\,mK stage (Fig.~\ref{fig:snout}). It is based on a 1024-pixel TES array\cite{Smith2021} from NASA/GSFC. Two columns of 32 pixels of this array are connected to their associated cold readout electronics\cite{Doriese2016,Reintsema2003} provided by NIST. Each pixel is read by a first stage Superconducting Quantum Interference Device (SQUID). These SQUIDs, referred as MUX SQUIDs, can be turned on and off, or addressed, via Flux Actuated Switches (FASs).

The Time-Division Multiplexing (TDM)\cite{Durkin2019} allows to read sequentially each row of all columns simultaneously. This snout allows us to perform a 2$\times$32 multiplexing which is sufficient for a validation of the X-IFU baseline readout electronics. The multiplexed signal of the first stage SQUIDs is amplified by an array of SQUIDs (SQUID AMP), situated on a cold PCB electronics (3\,K card) attached to the 3K plate.

Because TES and SQUID devices are highly sensitive to magnetic field, the snout is enclosed in a niobium shield, superconducting below 9\,K, that sets a very low uniform ambient magnetic field needed to run ground experiments on TES microcalorimeters\cite{Miniussi2019}. A field coil has been placed at the level of the TES array. It produces a magnetic field perpendicular to the TES array, set to null the remaining field at the detectors level along this direction. 

The characterisation of the thermal performances of the cryostat has been checked with the Goddard and NIST cold detection chain installed:
\begin{itemize}
    \item[-] the temperature stability is around 5\,$\mu$K\,rms, under optimisation;
    \item[-] the hold time for one ADR recharge is 15\,h at 55\,mK. It allows spectra with high enough counting statistics to determine the resolution to a few times 0.01\,eV;
    \item[-] the residual magnetic field at the focal plane level, inside the Nb shield, is around 1$\mu$T when the cryostat is enclosed in a $\mu$-metal shield during cooldown.
\end{itemize}
All these measurements are in good agreement with all values found during our cryostat full characterisation\cite{Betancourt2021}.

\section{PERFORMANCES OF THE DETECTION CHAIN IN ELSA}
\label{sec:perfo}

In order to evaluate the performance of the 50\,mK test bench cold detection chain, a warm readout chain also provided by Goddard and NIST electronics is used as a starting point.

\subsection{Goddard-NIST readout electronics}
\label{sec:current}

\begin{figure} [ht]
   \begin{center}
   \begin{tabular}{c c} 
   \includegraphics[height=6cm]{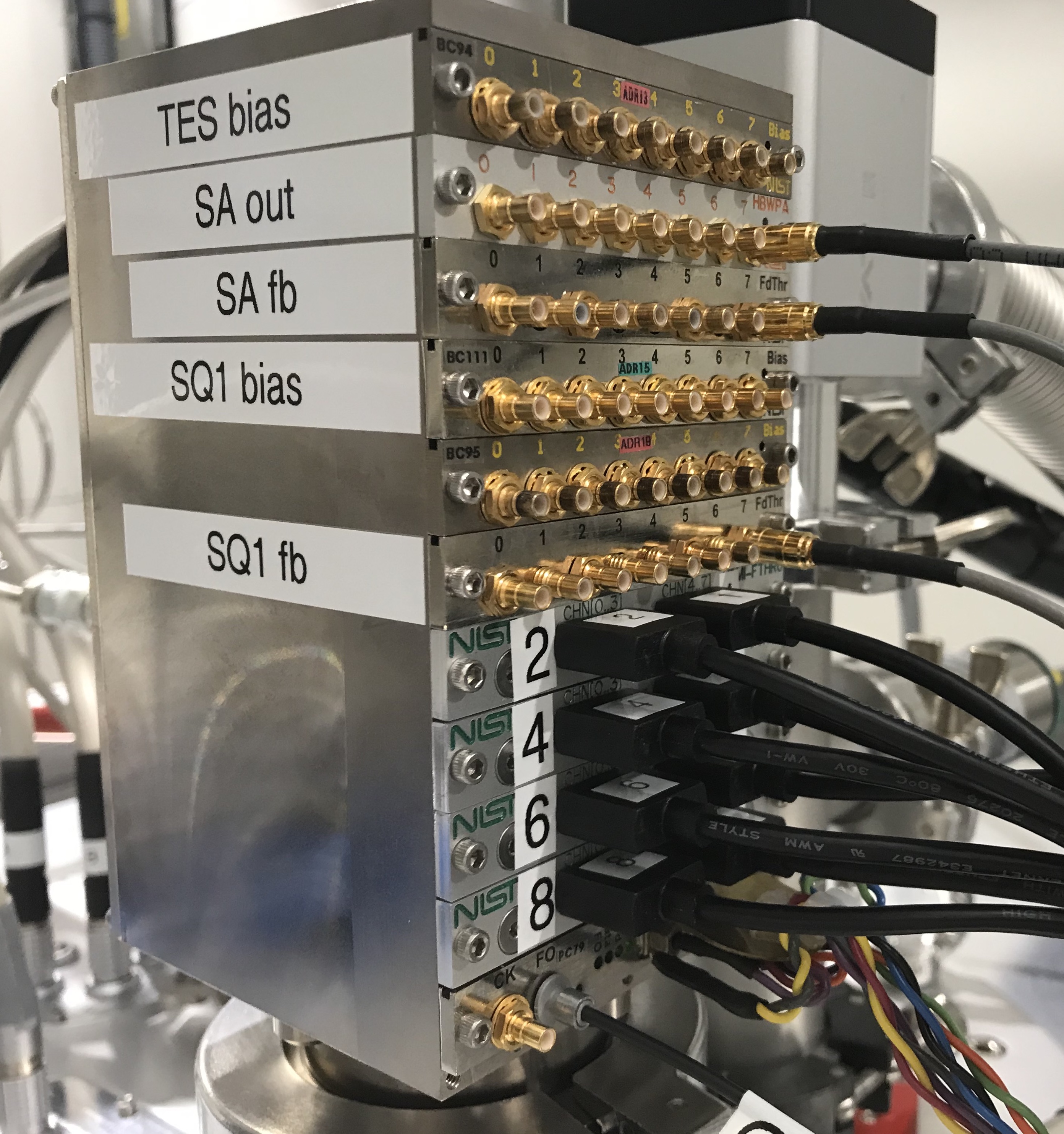} & \includegraphics[height=4.5cm]{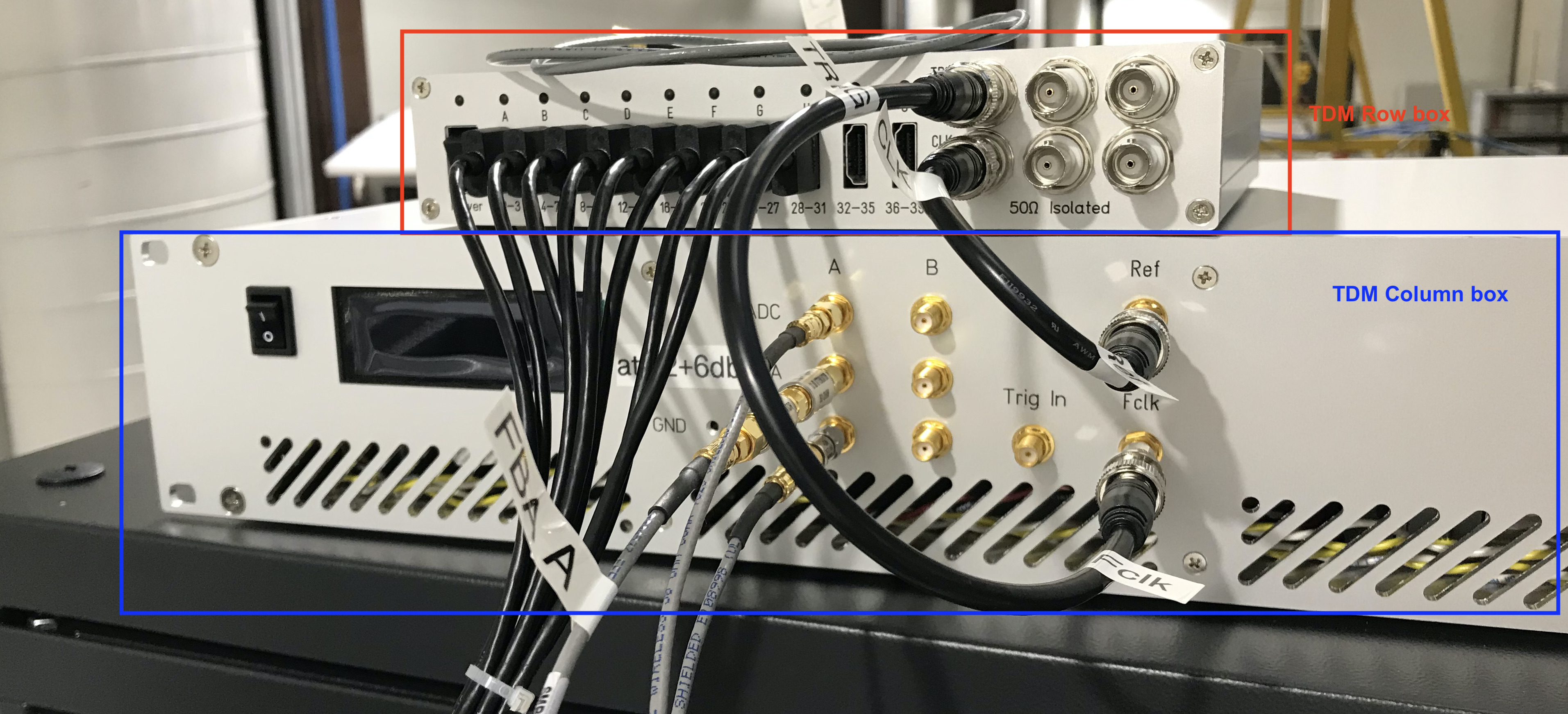}
	\end{tabular}
	\end{center}
   \caption[TDM] 
   { \label{fig:tdm} 
(Left) The Tower on top of the Elsa cryostat. SMB cards are labeled, and those cards allows to connect 8 columns. On the picture only one row is connected to TDM column box: SA out stands for output signal, SA fb for SQUID AMP flux bias and SQ1 fb for MUX SQUID feedback. 8 HDMI cables are connected to TDM row box for row sequencing on all columns simultaneously. (Right) The TDM readout electronics composed of the row box on top of the column box. The link between the two boxes are made with 2 BNC cables for clock synchronisation. The TDM column box has only one channel connected (A) on the picture but could allow 2 columns TDM acquisition.}
   \end{figure}

The “Tower”\cite{Reintsema2003} (NIST), on top of the cryostat (Fig.~\ref{fig:bench} and \ref{fig:tdm} left), is connected to the 3K card through a vacuum feedthrough by a set of flexes thermalized at 50\,K. It includes low-noise amplifiers for the TES signals, provides TES and SQUIDs biases, and feedthrough for feedback signals and for row addressing signals from the multiplexing electronics.

The TDM electronics\cite{Kazu2022} (NASA/GSFC) (Fig.~\ref{fig:bench} and \ref{fig:tdm} right) is made of commercial-off-the-shelf (COTS) electronic components. It is composed of the row electronics box which controls the row addressing by sending the sequencing signals to the FAS, via HDMI cables to the Tower. The row electronics box is connected via BNC cables to the column electronics box for clock synchronisation. The column electronics box de-multiplexes the TES signals, for up to 2 columns, and sends back the flux-locked loop (FLL) signals that allows the first stage SQUIDs to stay in a linear zone of their response.

\subsection{EMI/EMC reduction analysis}
\label{sec:EMI}

In order to perform a functional validation of the X-IFU instrument prototype warm readout chain, a satisfactory value of 3\,eV or better FWHM energy resolution at 5.9\,keV for a multiplexed acquisition has been chosen. Optimisation of the signal to noise ratio is performed through a careful EMI (Electro-Magnetic Interference)/EMC (Electro-Magnetic Compatibility) analysis and control implementation:
\begin{enumerate}
    \item As done on NIST and GSFC systems, a strict grounding scheme was implemented on the 50\,mK test bench:
    \begin{enumerate}
        \item All measurement electronics and cryostat control electronics are grounded through a single large copper braid;
        \item A metal cable path tray is installed between the cryostat structure and the measurement electronic rack.
    \end{enumerate}
    \item An isolation transformer allows to isolate the measurement electronics from any power supply disturbance.
    \item High-Frequency filtering is implemented at cryostat feedthroughs:
    \begin{enumerate}
        \item The magnet power-supply is filtered with Shaffner Single-stage Filter FN2410/FN2412;
        \item LEMO plugs for thermometry are filtered with 560\,pF capacitor EEseal filters performing a low-pass filter around 100\,kHz.
    \end{enumerate}
\end{enumerate}

Ground loops have been removed from the system, such as those created by the cable of the pressure gauge or the shielding of the field coil cable. These paths are removed permanently or just disconnected during experimentation. Further improvements are ongoing.

\subsection{Optical path alignment and filtering}
\label{sec:path}

In order to characterise our end-to-end present detection and readout electronics, a radioactive $^{55}$Fe source is placed in front of the down looking cryostat on-axis X-ray window. It emits X-ray photons in the Mn K$\alpha$ complex at 5.9\,keV. A remote controlled filter wheel selects a specific absorber (various Mylar and Al films) to obtain a typical count rate of 1\,photon/pixel/sec.

The optical path, allowing X-ray photons to reach the focal plane assembly inside the cryostat, is composed of a series of apertures at the bottom of each thermal shield. To block visible and infrared photons, each aperture has been filtered:
\begin{enumerate}
    \item[-] the Nb shield aperture filter is a 25\,$\mu$m Al foil;
    \item[-] both 3\,K and 50\,K aperture filters are aluminized mylar foils (6\,$\mu$m mylar coated with 200\,nm Al);
    \item[-] the 300\,K window, sustaining the differential pressure between vacuum and atmospheric pressure, is a Luxel ``LEX-HT" window providing a good X-ray transmission between 0.2 and 12\,keV.
\end{enumerate}
The alignment of the TES array center, the cryostat optical apertures, and the $^{55}$Fe source, has been checked. The deviation from the cryostat mechanical axis was determined to be smaller than 1\,mm for all optical items, excluding all possible vignetting.

\subsection{Test bench performance first validation}
\label{sec:test}

We present here the first validation of the full NASA/GSFC and NIST detection chain as implemented in the 50\,mK test bench, after performing the EMI/EMC optimisation detailed above. A dedicated integration (more than 6\,hours) on a 1$\times$8 pixels TDM subset setup allowed to detect at least 20.000 photons per pixel. Due to some false pulse triggers and multiple-pulse rejections, the number of final records were rather around 12.000 photons per pixel.

\begin{figure} [ht]
   \begin{center}
   \begin{tabular}{c} 
   \includegraphics[height=8.5cm]{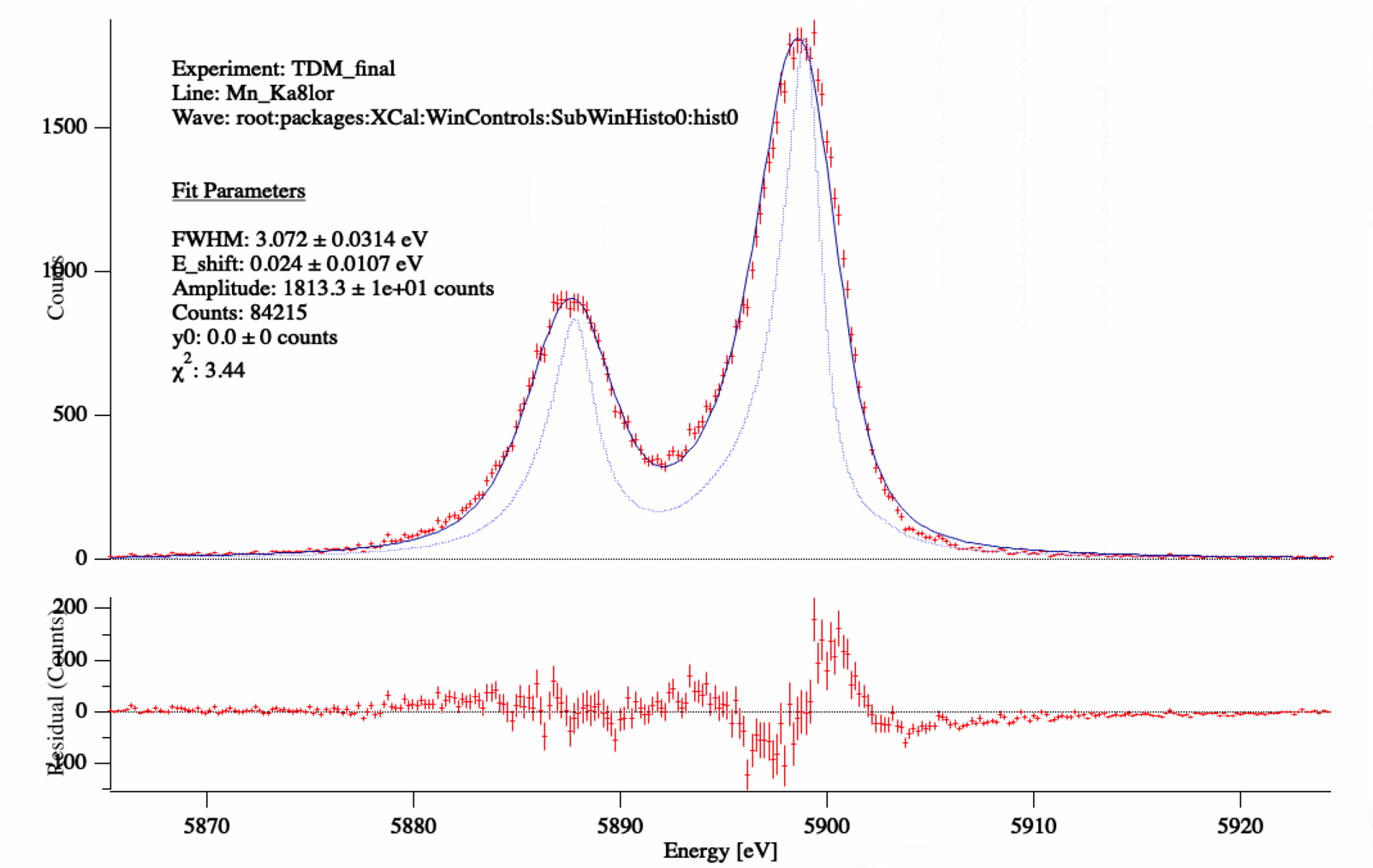}
   \end{tabular}
   \end{center}
   \caption[spectdm] 
   { \label{fig:spectdm} 
Mn-K$\alpha$ complex spectrum on 1$\times$8 multiplexed pixels after 6 hours of acquisition, processed with Goddard software.}
   \end{figure}

\begin{figure} [ht]
   \begin{center}
   \begin{tabular}{c} 
   \includegraphics[height=8.5cm]{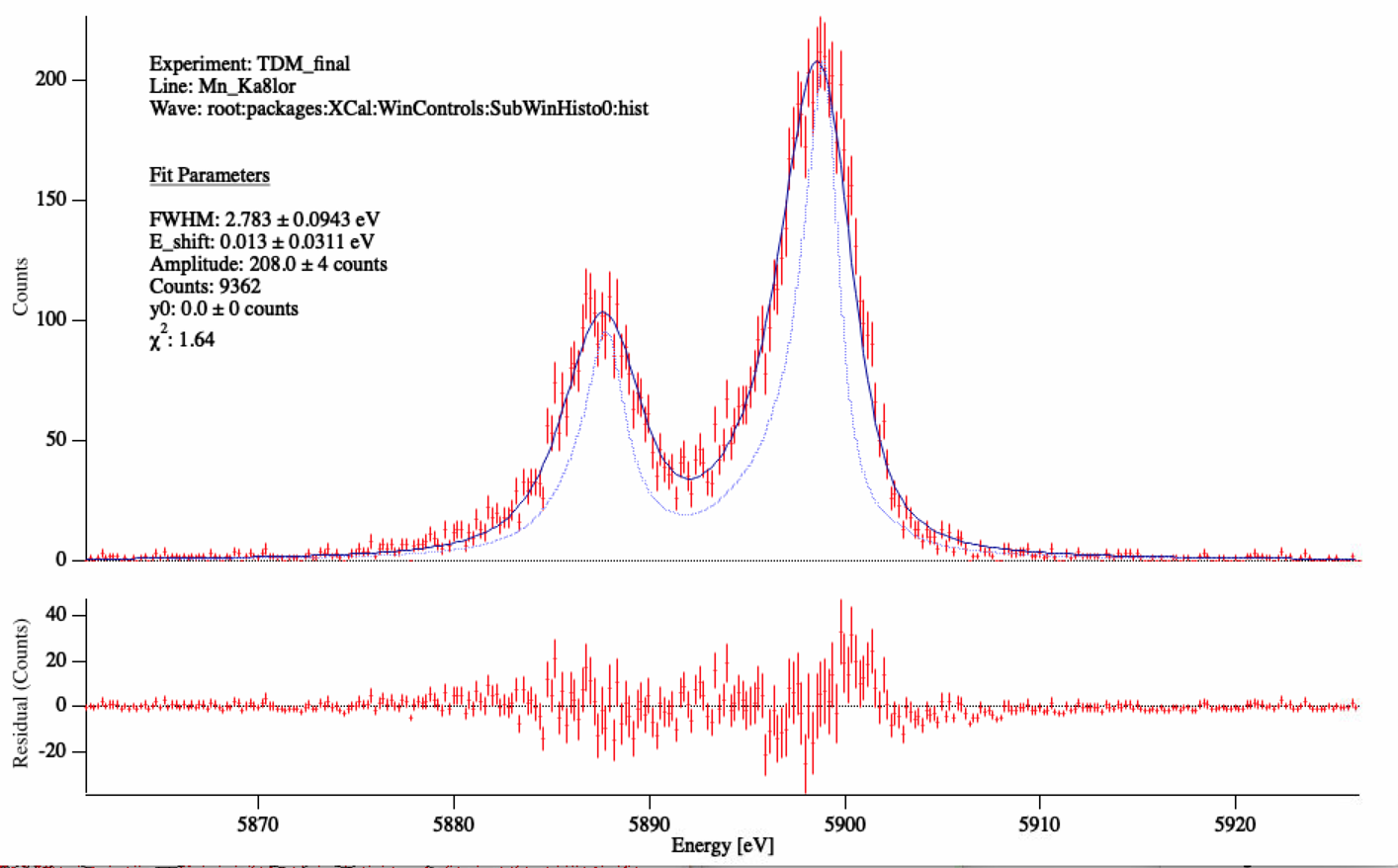}
   \end{tabular}
   \end{center}
   \caption[specsingle] 
   { \label{fig:specsingle} 
Same complex spectrum of Fig. \ref{fig:spectdm}, during the same acquisition, but for a single given pixel.}
   \end{figure}

In order to retrieve the photon energy thanks to pulse records, we apply \textit{optimal filtering} technique\cite{Fowler2016} using the NASA/GSFC X-ray data processing software, that implements all the protocols and algorithms needed. This analysis includes several processes such as the correction of the drift of the energy baseline during acquisition, the lag-phase correction and gain calibration. Spectra of the Mn K$\alpha$ complex were finally generated to assess the end-to-end energy resolution:
\begin{enumerate}
    \item[-] a 2.8\,eV FWHM energy resolution has been measured for a single given channel read in TDM (Fig.~\ref{fig:specsingle});
    \item[-] a 3.1\,eV FWHM energy resolution has been measured for the combined 8 TDM multiplexed pixels of one column (1$\times$8) (Fig.~\ref{fig:spectdm}).
\end{enumerate}

These results have to be compared to the performance of this detection chain in a GSFC cryostat: 2.1\,eV for a single channel and 2.6\,eV for a 2$\times$32 TDM acquisition\cite{Beaumont2022}. We assume that the difference might be due to still specific lines close to the sampling rate of our TDM frequency, visible on power spectrum density of our output baseline, together with the thermal stability of our system. We are presently performing tests to quantify the effect of bath temperatures variations on the energy resolution. We also observed that 2 of our 8 pixels have relatively lower resolution than others (3.4 and 3.5\,eV). We suppose that these might have reduced the overall energy resolution and caused those ``oscillations" on residuals in Fig. \ref{fig:spectdm}.

However, the energy resolution in the IRAP-CNES test bench is very close to the 3\,eV requirement in TDM for our demonstrations, and further EMI/EMC actions, as well as investigations to improve out thermal stability are in progress. This demonstrates the validation of the 50\,mK IRAP-CNES Elsa test bench for performing the end-to-end demonstration of the X-IFU prototype warm readout electronics.

\section{X-IFU PROTOTYPE WARM READOUT CHAIN VALIDATION}
\label{sec:validation}

The X-IFU baseline readout chain warm electronics is composed of:
\begin{enumerate}
    \item[-] the WFEE (Warm Front-End Electronics) is developed by APC\cite{Chen2018};
    \item[-] the DRE (Digital Readout Electronics) is developed by IRAP\cite{Ravera2018}.
\end{enumerate}

The purpose of the ELSA test bench is to perform a functional validation of the X-IFU baseline warm readout electronics in an end-to-end detection chain including microcalorimeters and a representative cold readout electronics. In order to perform this demonstration, prototypes of the WFEE and DRE will replace step-by-step the corresponding Goddard and NIST electronics previously used to validate the performance of the 50\,mK test bench.

\begin{figure} [ht]
   \begin{center}
   \begin{tabular}{c c} 
   \includegraphics[height=6cm]{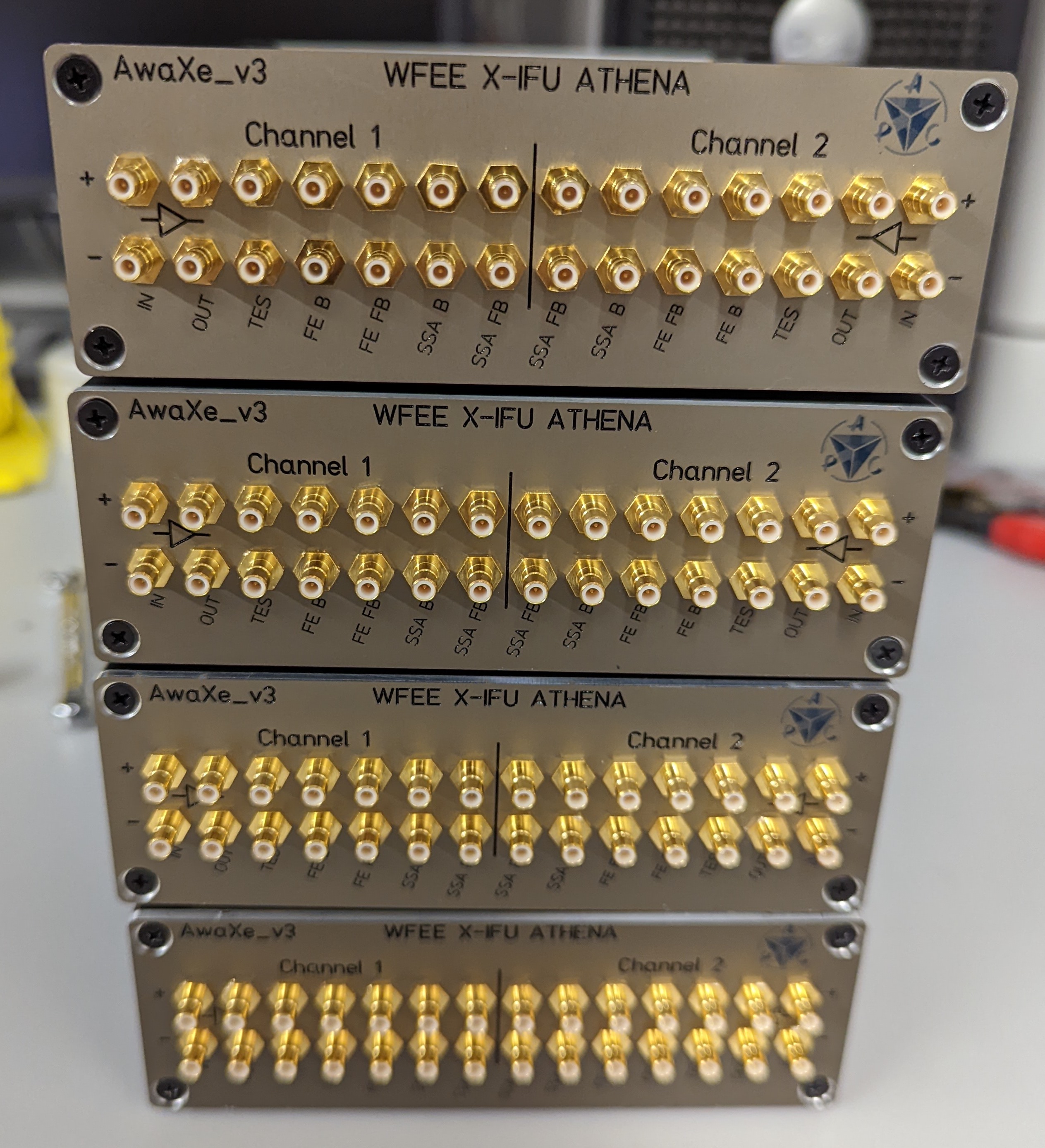} & \includegraphics[width=10cm]{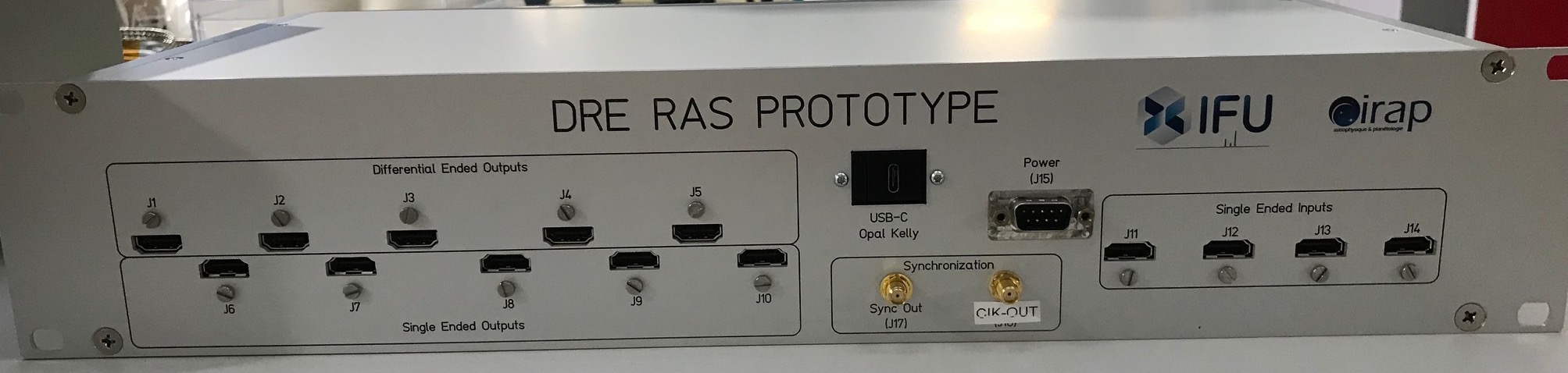}
	\end{tabular}
	\end{center}
   \caption[DRE] 
   { \label{fig:dre} 
(Left) Part of demonstration model of the Warm Front-End Electronics (WFEE). Each block are SMB connectors for signal output as well as TES bias, MUX SQUID bias and feedback and AMP SQUID flux bias inputs, for two channels (or columns). Those 4 blocks permits to perform 8 column TDM. (Right) Digital Readout Electronics Row Addressing and Synchronisation (DRE RAS) prototype module. There are 5 HDMI outputs (J1 to j5) for differential readout that will be used with WFEE. 5 other HDMI output (J6 to J10) are used in the actual single-ended readout electronics, allowing to test until 17 pixels in a column.}
   \end{figure}

The replacement of the Goddard TDM row electronics box by the prototype DRE Row Addressing and Synchronisation module (Fig.~\ref{fig:dre} right) is ongoing. First tests validate the compatibility between the RAS and the actual TDM Column box and Tower. Characteristic combined V-$\phi$ curves of the MUX SQUID and AMP SQUID, as well as X-ray photon pulses, have been measured using the DRE RAS prototype module, implementing the sequencing of a 1$\times$8 multiplexed acquisition.

The replacement of the TDM column box by a prototype DRE DEMUX (demultiplexing) model is planned next year, after  replacement of the NIST tower electronics by a demonstration model of the WFEE (Fig.~\ref{fig:dre} left). The WFFE and DRE electronics being differential electronics, the present single-ended 3\,K amplifying electronics and 3\,K-300\,K flexes will be upgraded using a differential electronics design.

All present analysis on the 50\,mK test bench are performed with a software suite from NASA/GSFC based on Igor Pro from Wavemetrics\footnote{\url{https://www.wavemetrics.com/}} (Fig.~\ref{fig:spectdm}). A software framework is developed by CNES: XIFUFWK. Based on the open-source Python language, XIFUFWK is designed to analyse data from the future X-IFU instrument. XIFUFWK is currently under validation on the 50\,mK test bench, using the NASA/GSFC software suite results in parallel.

\section{CONCLUSION}
\label{sec:conclusion}

We have performed a successful validation of a 50\,mK test bench to be used for the demonstration of the warm readout segment of the detection chain of the Athena/X-IFU. We measured a 3.1\,eV energy resolution at 5.9\,keV for a 1x8 multiplexing scheme using the warm readout from NASA/GSFC and NIST. Thermal stability and EMI/EMC planned further optimisations give us a reasonable insurance that our 3 eV energy goal resolution will be met. The X-IFU warm readout chain demonstration has started with a first flight compatible designed prototype for one component, the DRE RAS. Demonstration of the complete warm electronics segment can now be considered and planned with confidence.

\acknowledgments 
 
The material is partly based upon work supported by NASA under award number 80GSFC21M0002. 

\bibliography{report} 

\begin{thebibliography}{10}

\bibitem{Nandra2013}
{Nandra}, K. et~al., ``{The Hot and Energetic Universe: A White Paper
  presenting the science theme motivating the Athena+ mission},'' {\em arXiv
  e-prints} ,  arXiv:1306.2307 (June 2013).

\bibitem{Barriere2022}
{Barri{\`e}re}, N.~M. et~al., ``{Silicon Pore Optics},'' {\em arXiv e-prints} ,
   arXiv:2206.11291 (June 2022).

\bibitem{Meidinger2017}
{Meidinger}, N. et~al., ``{The Wide Field Imager instrument for Athena},'' in
  [{\em Society of Photo-Optical Instrumentation Engineers (SPIE) Conference
  Series}{\nolinebreak\hspace{0.1em}]},  {\em Society of Photo-Optical
  Instrumentation Engineers (SPIE) Conference Series} {\bf 10397},  103970V
  (Aug. 2017).

\bibitem{Barret2018}
{Barret}, D. et~al., ``{The ATHENA X-ray Integral Field Unit (X-IFU)},'' in
  [{\em Space Telescopes and Instrumentation 2018: Ultraviolet to Gamma
  Ray}{\nolinebreak\hspace{0.1em}]},  {den Herder}, J.-W.~A., {Nikzad}, S., and
  {Nakazawa}, K., eds., {\em Society of Photo-Optical Instrumentation Engineers
  (SPIE) Conference Series} {\bf 10699},  106991G (July 2018).

\bibitem{Pajot2018}
{Pajot}, F. et~al., ``{The Athena X-ray Integral Field Unit (X-IFU)},'' {\em
  Journal of Low Temperature Physics}~{\bf 193},  901--907 (Dec. 2018).

\bibitem{Doriese2017}
{Doriese}, W.~B. et~al., ``{A practical superconducting-microcalorimeter X-ray
  spectrometer for beamline and laboratory science},'' {\em Review of
  Scientific Instruments}~{\bf 88},  053108 (May 2017).

\bibitem{Smith2021}
{Smith}, S.~J. et~al., ``{Performance of a Broad-Band, High-Resolution,
  Transition-Edge Sensor Spectrometer for X-ray Astrophysics},'' {\em IEEE
  Transactions on Applied Superconductivity}~{\bf 31},  3061918 (Aug. 2021).

\bibitem{Doriese2016}
{Doriese}, W.~B. et~al., ``{Developments in Time-Division Multiplexing of X-ray
  Transition-Edge Sensors},'' {\em Journal of Low Temperature Physics}~{\bf
  184},  389--395 (July 2016).

\bibitem{Reintsema2003}
{Reintsema}, C.~D. et~al., ``{Prototype system for superconducting quantum
  interference device multiplexing of large-format transition-edge sensor
  arrays},'' {\em Review of Scientific Instruments}~{\bf 74},  4500--4508 (Oct.
  2003).

\bibitem{Durkin2019}
{Durkin}, M. et~al., ``{Demonstration of Athena X-IFU Compatible 40-Row
  Time-Division-Multiplexed Readout},'' {\em IEEE Transactions on Applied
  Superconductivity}~{\bf 29},  2904472 (Aug. 2019).

\bibitem{Miniussi2019}
{Miniussi}, A.~R. et~al., ``{Design of Magnetic Shielding and Field Coils for a
  TES X-ray Microcalorimeter Test Platform},'' {\em Journal of Low Temperature
  Physics}~{\bf 194},  433--442 (Mar. 2019).

\bibitem{Betancourt2021}
{Betancourt-Martinez}, G. et~al., ``{A test platform for the detection and
  readout chain for the Athena X-IFU},'' in [{\em Space Telescopes and
  Instrumentation 2020: Ultraviolet to Gamma Ray}{\nolinebreak\hspace{0.1em}]},
   den Herder, J.-W.~A., Nikzad, S., and Nakazawa, K., eds., {\em Society of
  Photo-Optical Instrumentation Engineers (SPIE) Conference Series} {\bf
  11444},  681 -- 691 (June 2021).

\bibitem{Kazu2022}
{Sakai}, K. et~al., ``{Developments of Laboratory-Based Transition-Edge Sensor
  Readout Electronics using Commercial-Off-The-Shelf Modules},'' in [{\em {19th
  International Workshop on Low Temperature Detectors
  (LTD19)}}{\nolinebreak\hspace{0.1em}]},  {\em {Journal of Low Temperature
  Physics}},  in press (July 2021).

\bibitem{Fowler2016}
{Fowler}, J.~W. et~al., ``{The Practice of Pulse Processing},'' {\em Journal of
  Low Temperature Physics}~{\bf 184},  374--381 (July 2016).

\bibitem{Beaumont2022}
{Beaumont}, S. et~al., ``{Development of an end-to-end demonstration readout
  chain for Athena/X-IFU},'' in [{\em {19th International Workshop on Low
  Temperature Detectors (LTD19)}}{\nolinebreak\hspace{0.1em}]},  {\em {Journal
  of Low Temperature Physics}},  in press (July 2021).

\bibitem{Chen2018}
{Chen}, S. et~al., ``{Development of the WFEE subsystem for the X-IFU
  instrument of the ATHENA Space Observatory},'' in [{\em Space Telescopes and
  Instrumentation 2018: Ultraviolet to Gamma Ray}{\nolinebreak\hspace{0.1em}]},
   {den Herder}, J.-W.~A., {Nikzad}, S., and {Nakazawa}, K., eds., {\em Society
  of Photo-Optical Instrumentation Engineers (SPIE) Conference Series} {\bf
  10699},  106994P (July 2018).

\bibitem{Ravera2018}
{Ravera}, L. et~al., ``{First results of the ATHENA/X-IFU digital readout
  electronics prototype},'' in [{\em Space Telescopes and Instrumentation 2018:
  Ultraviolet to Gamma Ray}{\nolinebreak\hspace{0.1em}]},  {den Herder},
  J.-W.~A., {Nikzad}, S., and {Nakazawa}, K., eds., {\em Society of
  Photo-Optical Instrumentation Engineers (SPIE) Conference Series} {\bf
  10699},  106994V (July 2018).

\end{thebibliography}
\bibliographystyle{spiebib} 

\end{document}